\documentclass[prd,preprint,tightenlines,floatfix,showpacs,preprintnumbers,nofootinbib,eqsecnum]{revtex4}
 \usepackage[dvips,final]{graphicx}
  \usepackage{amssymb}     \usepackage{amsmath}
    \usepackage{amsfonts}       \usepackage{epsfig}
      \usepackage{bm}
\def\Green{}
\def\Red  {}
\def\Black{}
\def\Blue {}
\newcommand\myBlu[1]{\textbf{\Blue #1}\Black}
\newcommand\myRed[1]{\textbf{\Red #1}\Black}
\newcommand\myGre[1]{\textbf{\Green #1}\Black}
\def\GeV{\text{{\rm GeV}}}  \def\MeV{\text{{\rm MeV}}}

\newcommand{\ac}{\mathcal{A}} 
\newcommand{\beq}{\begin{equation}} \newcommand{\eeq}{\end{equation}}

\def\al{\relax\ifmmode\alpha\else{$\alpha${ }}\fi}
\def\alps{\relax\ifmmode\alpha_s\else{$\alpha_s${ }}\fi}
\def\as{\relax\ifmmode\alpha_s\else{$\alpha_s${ }}\fi}
\def\msbar{\relax\ifmmode\overline{\rm MS}\else{$\overline{\rm MS}${ }}\fi}
\def\albar{\relax\ifmmode{\bar{\alpha}}\else{$\bar{\alpha}${ }}\fi}
\def\albars{\relax\ifmmode{\bar{\alpha}_s}\else{$\bar{\alpha}_s${ }}\fi}
\def\acal{\relax\ifmmode{\cal A}\else{${\cal A}${ }}\fi}
 \def\acalk{\relax\ifmmode{\cal A}_k\else{${\cal A}_k${ }}\fi}
 \def\acalkQ{\relax\ifmmode{\cal A}_k(Q^2)\else{${\cal A}_k(Q^2)${ }}\fi}
\def\alphaMs{\relax\ifmmode\alpha_M(s)\else{$\alpha_M(s)${ }}\fi}
\def\alphaEQ{\relax\ifmmode{\alpha_E(Q^2)}\else{$\alpha_E(Q^2)${ }}\fi}
\def\alphaM{\relax\ifmmode{\alpha}_M\else{$\alpha_M${ }}\fi}

 \graphicspath{./BAPTfigs/}
\begin{document}
\thispagestyle{empty} \preprint{\hbox{}} \vspace*{-10mm}

\title{Bjorken Sum Rule and pQCD frontier on the move}
\author{Roman~S.~Pasechnik}
\author{Dmitry~V.~Shirkov}
\author{Oleg~V.~Teryaev}

\affiliation{Bogoliubov Lab, JINR, Dubna 141980, Russia}
\date{\today}

\begin{abstract}
 The reasonableness of the use of perturbative QCD notions in the
 region close to the scale of hadronization, i.e., below $\lesssim 1 \GeV$
 is under study. First, the interplay between higher orders of pQCD
 expansion and higher twist contributions in the analysis of recent
 Jefferson Lab (JLab) data on the Generalized Bjorken Sum Rule function
 $\Gamma_1^{p-n} (Q^2)$ at $0.1<Q^2< 3\,{\rm GeV}^2$ is
 studied. It is shown that the inclusion of the higher-order pQCD
 corrections could be absorbed, with good numerical accuracy, by
 change of the normalization of the higher-twist terms.
  Second, to avoid the issue of unphysical singularity (Landau pole
 at $Q=\Lambda\sim 400\,\MeV\,$), we deal with the ghost-free
 Analytic Perturbation Theory (APT) that recently proved to be
 an intriguing candidate for a quantitative description of light
 quarkonia spectra within the Bethe-Salpeter approach.
The values of the twist coefficients $\mu_{2k}\,$ extracted from the
 mentioned data by using the APT approach provide a better convergence of
 the higher-twist series than with the common pQCD. As the main result, a
 good quantitative description of the JLab data down to $Q\simeq$ 350 MeV
 is achieved.
\end{abstract}

\maketitle
\section{Introduction}

  The analysis of Deep Inelastic Scattering (DIS) data by
 combination of Perturbative Quantum Chromodynamics (pQCD) and
 Operator Product Expansion (OPE) provides us with a test site
 for combining both the perturbative and non-perturbative (NP) QCD
 contributions in the low energy domain. In particular, the
 Generalized ($Q^2$-dependent) Bjorken Sum Rule (BSR)~\cite{Bj66}
 is a renown target ground for testing different possibilities
 \cite{Kodaira79,SofTer}. Fortunately, fresh Jefferson Lab data
 \cite{Deur:2008rf} give information on the spin-dependent BSR
 amplitude $\Gamma_1^{p-n}(Q^2)\,$ behavior close to the
 confinement/hadronization scale. Meanwhile, in this region the
 common theoretical pQCD analysis is spoiled by the unphysical
 singularities in the infrared (IR) region at a scale
 $\sim \Lambda\sim 400\,\MeV\,.$ \\
  To cure this disease (known as the Landau pole trouble) of the
 pQCD expansion, we use the Analytic Perturbation Theory (APT)
 approach \cite{apt96-7} based on the causality principle
 implemented as analyticity imperative in the complex
 $Q^2$-plane for the QCD coupling $\alpha_s(Q^2)\,$ in the form of
 the K\"allen-Lehmann spectral representation (for a review on
 APT concepts and algorithm see Ref.~\cite{Sh-revs}).

 In principle, the  shift of the pQCD frontier (i.e., the boundary
 above which pQCD is applicable) and the rearrangement of
 the total contribution between perturbative and non-perturbative
 terms is possible by an appropriate modification of perturbative
 series. Examples are provided, say, by IR renormalons
 \cite{renormalon,kataev} and the extractions of higher twists (HT)
 using various approximations of pQCD \cite{SidKat}. In the present
 paper, we systematically explore the possibility of such a shift by
 extracting the values of HT terms using various approximations and
  modifications of PT
 to analyse recent JLab data \cite{CLAS,Deur,chen06} on the BSR.

 We found that a particular form of solution to the Renormalization
 Group (RG) equation, namely, the ``denominator'' form~\cite{denom06}
 (see Eq.~(\ref{Denom}) below) is much more suitable for the use in
 the low-$Q^2$ region than the most popular ones based on the
 ``multistory'' Eq.~(9.5) of the Particle Data Group~\cite{pdg06}
 compendium (Eq.~(\ref{9.5pdg}) below). The inclusion of
 higher-order (HO) of PT demonstrates a ``duality'' between HO and HT,
 in the sense that HT terms are absorbed, with good numerical accuracy,
 into HO terms. As a result, HT coefficients decrease. At the same time,
 we observed that the description of the data is improved only up to
 the two-loop order of PT, which may be a signal of the asymptotic
 character of PT series in the region close to the Landau pole.

  A further shift of the pQCD frontier is achieved by using the analytic
 APT modifications of pQCD in the analysis of BSR below 1 GeV making
 possible investigation of the interplay of  PT and HT contributions.
 As a result, we find that while the next-to-leading twist ($\mu_4$)
 term is larger in APT than in the usual pQCD, the HT ($\mu_{6,8}$)
 coefficients in APT are smaller (making the impression of HT series
 being convergent) and allowing for a reasonable description of the data
 down to $Q \sim 350$ MeV.

\section{The Bjorken Sum Rule in conventional PT}

The Bjorken integral is defined
\begin{eqnarray}\label{eq1}
\Gamma_1^{p-n}(Q^2)=\int^1_0dx(g^p_1(x,Q^2)-g^n_1(x,Q^2))\,,
\end{eqnarray}
 via the spin-dependent proton and neutron structure functions
 $g^{p}_1\,,\,g^{n}_1\,$  with $x=Q^2/2M\nu$, the energy transfer $\nu$
 and the nucleon mass $M.$ At large $Q^2$, the BSR comes to its renowned
 form $\Gamma_1^{p-n}=g_A/6,$ where $g_A=1.267\pm0.004$ is the nucleon
 axial charge defined from the neutron $\beta$-decay. At finite
 $Q\gg\Lambda\,,$ the BSR is given by the OPE series in $1/Q^{i-2}$ with
 even $i=2,4\dots$ being the number of a twist and the pQCD series in
 $\alpha_s^n\,.$ The expression for the perturbative part of
 $\Gamma^{p-n}_{1}(Q^2)$ including the HT contribution is (see e.g.
 Ref.~\cite{kataev})
 \begin{eqnarray}\nonumber
 \Gamma^{p-n}_{1,PT}(Q^2)=\frac{g_A}{6}\biggl[1-
 \frac{\alpha_s}{\pi}- 3.558\left(\frac{\alps}{\pi}\right)^2 \\
 -20.215\left(\frac{\alps}{\pi}\right)^3 -O(\alps^4)
 \biggr]+\sum_{i=2}^{\infty}\frac{\mu_{2i}}{Q^{2i-2}}\label{PT-Bj}
 \end{eqnarray}
 with numerical values given at $n_f=3$ and weak dependence of
 $\mu_{2i}$ on $\log Q^2$ neglected.
 The first non-leading twist
 term \cite{Shuryak} can be expressed \cite{chen06}
 \begin{eqnarray}\nonumber
 \mu_4^{p-n}\approx\tfrac{4\,M^2}{9}f_2^{p-n},
 \end{eqnarray}
  in terms of the colour polarizability $f_2$ .

 Within the pQCD, the $\as$ coupling is usually taken in the form
 {\small
 (Eq.~(7) in~\cite{bethke}, Eq.~(9.5) in PDG~\cite{pdg06})}
 expanded over $\ln L/L\,$ with ($L=\ln(Q^2/\Lambda^2),\,
 b_k =\beta_k/\beta_0$)
 $$
 \albars^{(4)}(L)=\tfrac{1}{\beta_0L}-\myBlu{$\tfrac{b_1}{\beta_0^2}
 \tfrac{\ln L}{L^2}$}+
 \framebox{$\myBlu{$\tfrac{1}{\beta_0^3L^3}$}
 \left[\myBlu{$b_1^2(\ln^2L-\ln L-1)$}+\myRed{$b_2$}\right]$} $$
 \vspace{-8mm}

  {\small
 \begin{eqnarray}\label{9.5pdg}
 -{\tfrac{1}{\beta_0^4L^4}}\left[\myBlu{$ b_1^3
 \left(\ln^3L -\tfrac{5}{2}\ln^2L-2\ln L+\tfrac{1}{2}\right)$}+
 \myRed{$3b_1b_2\ln L$}-\myGre{$\tfrac{b_3}{2}$}\right].\,
  \end{eqnarray} }  \vspace{-5mm}

  Here, the second term in the first line is the 2-loop contribution and
 the framed term usually is referred to as ``the 3-loop one'', while all
 contents of the second line is treated on the equal footing with the
 4-loop term \myGre{$\sim b_3\,.$}

  However, it is evident that pieces of genuine 2-loop contribution
 proportional to $b_1\,$ are entangled with the higher-loop
 ones. This defect is absent in the more compact {\sf``Denominator
 representation"}~\cite{denom06},
  {\small  \begin{equation}\label{Denom}
 \tfrac{1}{\albars^{(\myGre{$3$})D}(L)}= \beta_0\,L+ \myBlu{$b_1$}
 \left[\myBlu{$\ln L\,+\ln\left(1+\tfrac{b_1\,\ln L}{\beta_0\,L}\right)$}
 +\tfrac{\myBlu{$b_1^2$}-\myRed{$b_2$}}{\beta_0\,L}\right]\end{equation}}
 which, being generic for the PDG expression, is closer to the iterative
 RG solution and, hence, more precise. Below, we shall refer to it as to
 ``Denom''.

 A detailed higher-twist analysis based on the total set of low energy
 SLAC and JLab data was performed in Ref.~\cite{Deur}. The result of
 the combined fit done in the $Q^2$-range 0.66-10.0 $\GeV^2$ is
 {\small$f_2(Q^2=1\,{\rm GeV}^2)=-0.101\pm 0.027$} and
 {\small $\mu_6/M^4 =0.084\pm 0.011$} (elastic contribution included).
  The fitting procedure of Refs.~\cite{Deur,chen06} taking the pQCD
 leading-twist term calculated at NLO $\alpha_s$ and using the 2-loop
 ``Denom'' coupling $\bar\alpha_s^{(2),D}$ was repeated. We
  succeeded in obtaining the central values
 of Refs.~\cite{Deur,chen06} in the two-parametric fit with the output
  {\small $$
 f_2=-0.096\pm 0.012,\;\,\mu_6/M^4 =0.087\pm 0.004,\,\;\chi^2=0.48, $$}
 where the errors are statistical only. These results are compatible
 with HT extraction performed in Ref.~\cite{Stamenov}.

 It is of special interest to study the BSR data with the elastic
 contribution (necessarily present in the OPE framework \cite{Ji})
 excluded, since the low-$Q^2$ behavior of such an ``inelastic'' BSR
 integral (coinciding with the usual BSR for $Q^2\to\infty$) is
 constrained by the Gerasimov-Drell-Hearn (GDH) sum rule \cite{GDH}, and
 one may investigate its continuation to low energies \cite{SofTer}.
 To this goal, doing the same NLO combined fit (elastic contribution
 excluded), one gets
 {\small $$f_2^{\rm inel}=-0.080\pm 0.016,\,\;\mu_6/M^4=0.022\pm 0.005,\;
 \chi^2=0.91\,. $$}
 The difference is noticeable starting from $\mu_6$ which is natural
 due to a decrease of an elastic contribution with growing $Q^2$.

 To explore the fit results sensitivity to the PT order and to
 the form of $\alpha_s$ below 1 GeV, we have also performed
 fits at the 1-, 2- and 3-loop levels. The minimal borders of fitting
 domains in $Q^2$ were settled from the {\it ad hoc} restriction
 $\chi^2\leqslant1$.

\begin{figure}[h!]
 \centerline{\epsfig{file=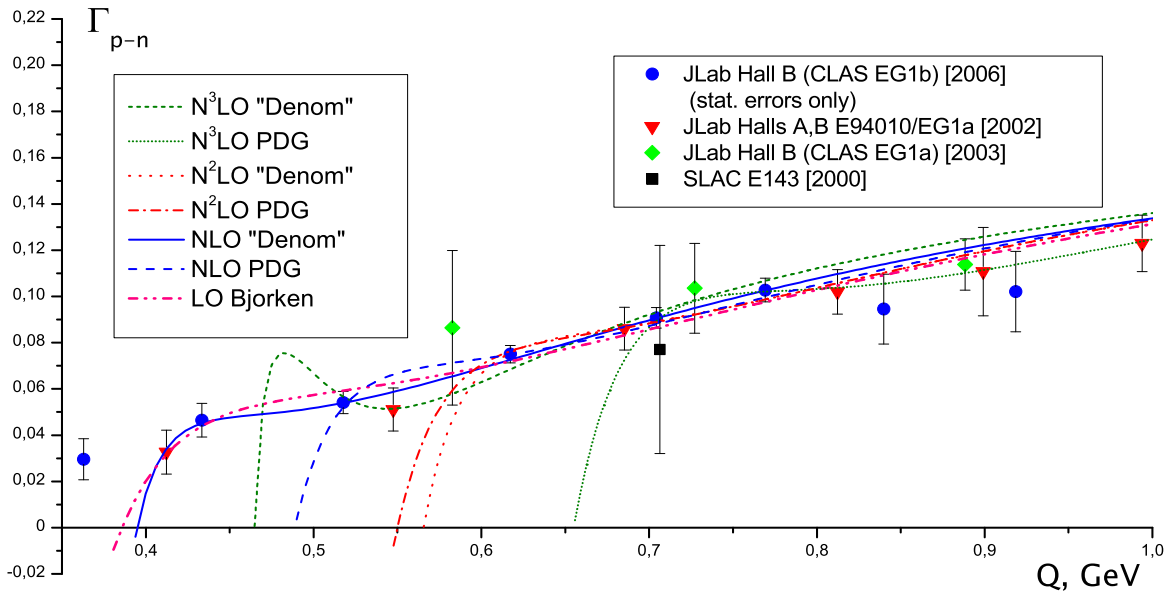,height=8cm,width=15.5cm}}
 \caption{\footnotesize Best 3-parametric fits of JLab and SLAC data
 on Bjorken SR calculated within  $\alpha_s$ in PDG form (\ref{9.5pdg})
 and in the ``Denom'' one (\ref{Denom}) at various loop orders.}
 \label{fig:loop-sen}  \end{figure}

  From Fig.~\ref{fig:loop-sen}, one sees that the results obtained with
 2- and 4-loop expressions for the ``Denom''
 coupling are better consistent with the BSR data at $Q < 1$ GeV than
 those based on the PDG expression (\ref{9.5pdg}), though 3-loop
 results do not differ significantly.

  Indeed, by using Eq.~(\ref{Denom}) one may extend the applicability
 domain of Eq.~(\ref{PT-Bj}) down to $Q^2\sim 0.27$ GeV$^2$. At the same
 time, Eq.~(\ref{9.5pdg}) works well only down to $\sim 0.47$ GeV$^2$
 due to extra $\ln^nL/L^{n+1}$ singularities.

\begin{figure}[h!]
 \centerline{\epsfig{file=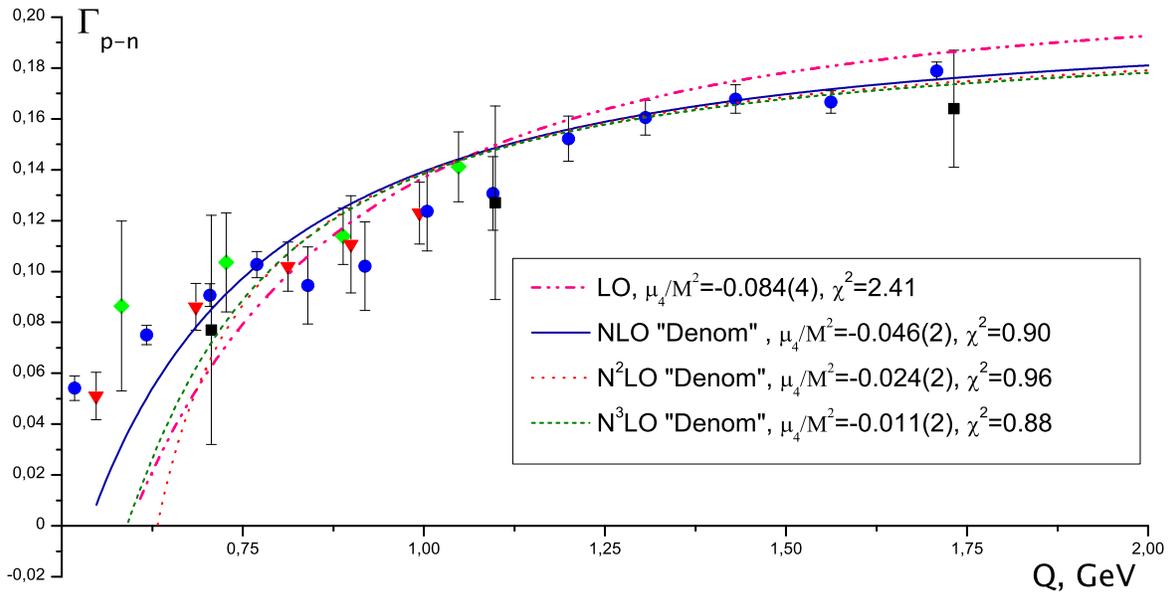,height=8cm,width=15.5cm}}
 \caption{\footnotesize One-parametric fits of the JLab and SLAC data
 on Bjorken SR calculated in the ``Denom'' form in different loop orders.}
 \label{fig:dual}
\end{figure}

 The fitting of BSR data in 2,3,4-loops over the fixed range
 $0.6\,\GeV<Q< 2.0\,\GeV\,$
 yields  a ``swap" between the higher orders of PT and HT terms. In
 Fig.~\ref{fig:dual}, we show one-parametric fits with 2,3-loop $\as$
 pQCD to the BSR. One can see there that the higher-loop contributions
 are effectively ``absorbed" into the value of $\mu_4$ which
 magnitude decreases with increasing of the loop order while all the
 fitting curves are very close to each other. This observation reveals
 a kind of "duality" between perturbative $\as$-series
 and nonperturbative $1/Q^2$-series.

  This also means the appearance of a new aspect of quark hadron duality,
 the latter being the necessary ingredient of all the QCD applications
 in the low energy domain. Usually, it is assumed \cite{Shifman:1978bx}
 that the perturbative effects are less important there than the power
 ones due to a nontrivial structure in QCD vacuum.

 In our case, the PT corrections essentially enter into the
 game, so that the pQCD HO terms are relevant in the domain where the
 concepts of  traditional hadronic physics are usually applied.

 The interplay between partonic and hadronic degrees of freedom in the
 description of GDH SR and BSR  may be also observed in the surprising
 similarity between the results of ``resonance'' \cite{Burkert:1992tg}
 and ``parton'' \cite{SofTer} approaches.

 At the same time, from
 Fig.~\ref{fig:loop-sen} it follows that the higher (3- and 4-loop) PT
 orders yield a worse description of the BSR data, probably
 implying the asymptotic character of the series in powers of $\as$.

 One may ask to what extent the troubles mentioned above are due to
 the unphysical singularities at $Q\sim \Lambda$ in PT series for
 $\Gamma^{p-n}_{1, PT}$. Their influence is essential just at $Q< 1$
  GeV where the HT terms play an important role.

 The APT is free of such problems, thus providing a tool to
 investigate the behavior of  HT terms extracted directly from the low
 energy data. This provides a motivation for the analysis performed in
 the next section.

 \section{The Bjorken Sum Rule in APT}
 According to the approach developed by Igor Solovtsov and co-authors
 \cite{MSS_Bj:98}, the APT modification of BSR with HT power
 corrections looks like
\begin{eqnarray}\nonumber
\Gamma^{p-n}_{1}(Q^2)&=&\Gamma^{p-n}_{1,APT}(Q^2)+
\sum_{i=2}^{\infty}\frac{\mu^{APT}_{2i}}{Q^{2i-2}},\\
\Gamma^{p-n}_{1,APT}(Q^2)&=&\frac{g_A}{6}\left[1-
\Delta^{p-n}_{\rm 1,APT}(Q^2)\right]\,,\label{APT-part}\end{eqnarray}
 {\small $$\Delta^{p-n}_{\rm 1,APT}=0.318\,{\cal A}^{(3)}_1(Q^2)+
  0.361\,{\cal A}^{(3)}_2(Q^2)
  +0.652{\cal A}^{(3)}_3(Q^2)+... $$}

 It should be noted that the APT Euclidean functions
 in the 1-loop case are simple enough \cite{apt96-7}
{\small\begin{eqnarray}                                    
 &&\acal_1^{(1)}(Q^2)=\frac{1}{\beta_0}\left[\frac{1}{L}+
 \frac{\Lambda^2}{\Lambda^2-Q^2}\right]\,,\quad
 L=\ln\left(\frac{Q^2}{\Lambda^2}\right),\label{AE1-2}\\
 &&\acal_2^{(1)}(l)=\frac{1}{\beta_0^2}\left(\frac{1}{L^2}-
 \frac{Q^2\,\Lambda^2}{(Q^2-\Lambda^2)^2}\right),\;
 \acal_{k+1}^{(1)}=-\,\frac{1}{k\,\beta_0}\,
 \frac{d\,\acal_k^{(1)}} {d L} \,.\nonumber
\end{eqnarray}}
 \noindent
 the higher ${\cal A}_k$ being related to the lower ones recursively
 by differentiating. Analogous 2- and 3-loop level expressions involve
 a little known special Lambert function and are more intricate
 ~\cite{Magr:00,K-Magr:01}.

 Meanwhile, even for the 3-loop APT case, there exists a possibility
 to employ {\it the effective log approach} proposed by Igor Solovtsov
 and one of the authors~\cite{SolSh99} and extended recently (see
 Eqs.~(12), (14) in Ref.~\cite{ShZ05}) to higher APT Euclidean and
 Minkowskian functions. In the present context, one may use simple model
 one-loop expressions (\ref{AE1-2}) with some {\it effective 2-loop
 log} $L^*\,$ accumulating the 2-loop log-of-log
 {\small
 \begin{eqnarray} \label{model2}                       
 &&\acal_{1,2,3}^{(3)}(L)\to\acal_{1,2,3}^{mod}=\,\acal_{1,2,3}^{(1)}(L^*)
 \,;\\ &&L^*= L+B(n_f)\ln\sqrt{L^2+2\pi^2}\,;
 \qquad B=\beta_1/\beta_0^2\,.
 \nonumber  \end{eqnarray}}
 Happily enough, the second term does not undergo a significant
 variation in the intermediate few GeV region. Indeed, as
 $B(n_f=3)=0.79\,$ and $B(4)=0.74\,,$ in the 
 region $\,Q<5\,\,\GeV\,$ involved in the BSR data analysis, a simple
 approximation
{\small\begin{eqnarray}\label{1parmodel-mod}                    
 &&L^*\simeq L+B(n_f)\ln\sqrt{2\pi^2}=2\,\ln(Q/\Lambda^{(1)}_{eff})\, ,\\
 &&\Lambda^{(1)}_{eff}=e^{-\frac12 B(n_f)\ln\sqrt{2\pi^2}}
 \Lambda^{(3)}\sim 0.50\,\Lambda^{(3)}\,\nonumber  \end{eqnarray}}
  happens to be accurate enough.
 That is, instead of the cumbersome 3-loop expressions for the APT
 functions, in Eq.~(\ref{APT-part}) one can use the 1-loop expressions
 (\ref{AE1-2}) with $\Lambda_{mod}=\Lambda^{(1)}_{eff}\,$ value given by
 the second relation (\ref{1parmodel-mod}).

 If we take $\Lambda^{(3)}=380\,\MeV$ \cite{Pro}, then
 $\Lambda^{(1)}_{eff}\simeq190\,\MeV$. The corresponding maximal errors
 of the model (\ref{model2}) for first and second functions are
 \cite{ShZ05} $\delta\ac^{mod}_1/\ac^{mod}_1\simeq4\%$ and
 $\delta\ac^{mod}_2/\ac^{mod}_2\simeq8\%$ at $Q\sim\Lambda^{(3)}\,,$
 which seems to be sufficiently accurate. Indeed, as far as
 $\acal_1(Q=400\,\MeV)=0.532\,$ and $\acal_2(400\,\MeV)=0.118\,,$
 the total error in $\Gamma^{p-n}_{\rm 1,APT}\,$ is mainly determined
 by the first term, being of the order $\delta\Gamma^{p-n}/\Gamma^{p-n}
 \simeq\delta\ac^{mod}_1/\pi\sim1\,\%\,,$ i.e., less than the data
 uncertainty.

 Turn now to the 3-loop APT part of the Bjorken integral
 $\Gamma^{p-n}_{1,APT}(Q^2)$. Its value is quite stable with respect to
 small variations of $\Lambda\,$ (in contrast with huge instability of
 $\Gamma^{p-n}_{1,PT}\,$): it changes now by about $2-3\%$ within the
 interval  $\Lambda^{(3)}=300-400\,\MeV\,$
 \footnote{In particular, this means that the low-$Q$ BSR data cannot
  be used for a determination of $\Lambda$ in the APT approach.}.

\begin{figure}[h!]
 \centerline{\epsfig{file=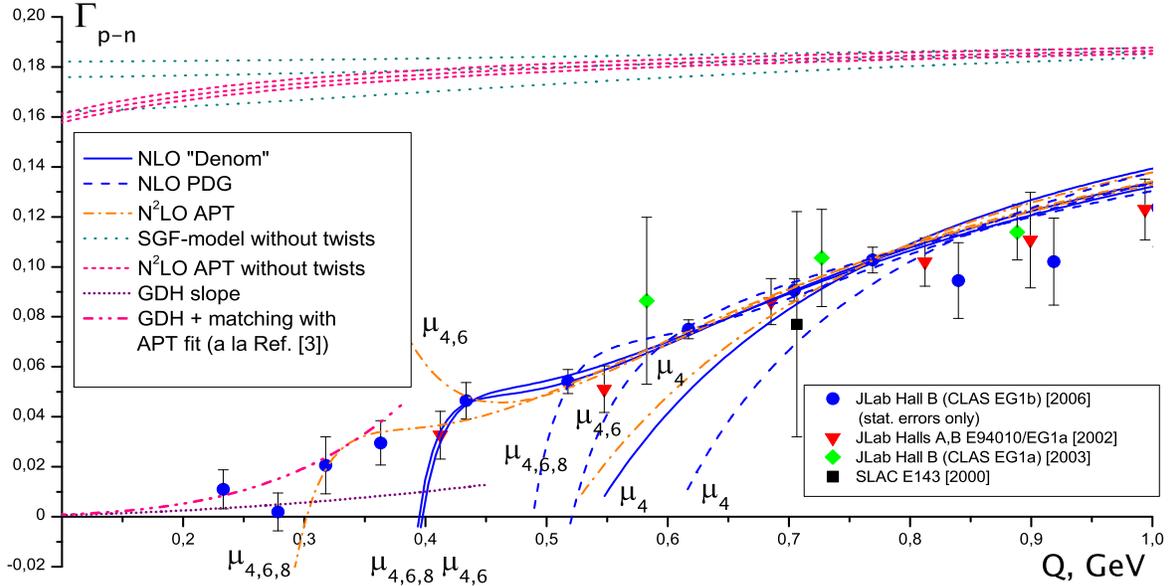,height=8cm,width=15.5cm}}
 \caption{\footnotesize Best 1,2,3-parametric fits of the JLab and SLAC
 data on Bjorken SR calculated with NLO "Denom" (solid lines) and PDG
 (dashed lines) couplings and N$^2$LO APT (dash-dotted lines) at
 fixed $\Lambda_{QCD}$ value corresponding to the world average. We
 also show the pQCD part of the BSR at different values of
 $\Lambda^{(3)}=300,\,350,\,400$ MeV calculated within APT (short-dashed
 lines) and SGF-model \cite{Simonov} at different values of the glueball
 mass $M_0=1.2,\,1.0,\,0.8\,\GeV$ (with $\Lambda=380$ MeV) (dotted
 lines).} \label{fig:tot}\end{figure}

 We also performed the comparison with Simonov's ``glueball-freezing
 model'' (SGF-model) \cite{Simonov} -- see Fig.~\ref{fig:tot}, with
 similar to PDG $1/L$-type loop expansion for freezed coupling
{\small\begin{eqnarray}\label{simon} \nonumber
 \phantom{AAAAAA}\alpha_B(Q^2)=\alpha_s^{(2)}(\bar{L})\,,\quad
 \bar{L}=\ln(\tfrac{Q^2+M_0^2}{\Lambda^2})\qquad\quad\mbox{(SGF)}
 \end{eqnarray}}
 where 2-loop $\alpha_s^{(2)}$ is taken in the form of two terms from
 the first line in Eq.~(\ref{9.5pdg}) with logarithm modified by a
 ``glueball mass'' $M_0\sim 1\, \GeV$ and the usual PT expansion in
 powers of $\alpha_B$ in $\Gamma^{p-n}$ is adopted.

 Extending the analysis of Ref.~\cite{MSS_Bj:98} to lower $Q$ values, we
 estimated the relative size of APT contributions to the BSR. It turned
 out that the third term $\sim\ac_3$ contributes no more than $5\%$ to
 the sum, thus supporting the practical convergence of the APT series.

  Note that the APT functions $\ac_k$ contain the $(Q^2)^{-k}\,$ power
 contributions which effectively change the fitted values of $\mu$-terms.
 In particular, subtracting extra $(Q^2)^{-1}$ term induced by the APT series
 {\small\begin{eqnarray*}
 &&\Gamma^{p-n}_{1,APT}(Q^2)\simeq\frac{g_A}{6}+
 f\biggl(\tfrac{1}{\ln(Q^2/{\Lambda^{(1)}_{eff}}^2)}\biggr)+
 \varkappa\tfrac{{\Lambda^{(1)}_{eff}}^2}{Q^2}+ {\cal O}(\tfrac{1}{Q^4})
 \end{eqnarray*}}
 with $\varkappa=0.43$, we get
{\small\begin{eqnarray}                                
\label{mu4_APT}
 \frac{\mu_4^{APT}+\varkappa{\Lambda^{(1)}_{eff}}^2}{M^2}\simeq
 \frac{\mu_4}{M^2}\simeq-0.048\,,\quad \Lambda^{(1)}_{eff}\sim0.2\,\GeV\,,
 \end{eqnarray}}
 that nicely correlates with the result in Ref.~\cite{Deur}:
 $\mu_4/M^2\simeq-0.045.$ This demonstrates the concert of the APT
 analysis with the usual PT one for the BSR data at $Q^2\geq1\;\GeV^2$.

  In Fig.~\ref{fig:tot}, we show best fits of the combined data set for
 the function $\Gamma_1^{p-n}(Q^2)$ in the PT and the APT approaches.
  The corresponding numerical results are given in Table~\ref{table}.
 Our fit gives the HT values indicating a better convergence
 of the OPE series due to decreasing magnitudes and alternating signs
 of consecutive terms, in contrast to the usual PT fit results.
\begin{table}[h]
\caption{\small\sf Combined fit results for the HT terms in APT and
conventional PT in PDG and "Denominator" forms.}
\begin{center}\label{table}
\begin{tabular}{|c|c|c|c|c|} \hline
 $\quad$Method$\quad$& $Q_{min}^2,\,\GeV^2$ & $\mu_4/M^2$ & $\mu_6/M^4$ & $\mu_8/M^6$ \\ \hline\hline
 NLO PDG             &  0.50 & -0.043(2)  &     0      &    0       \\ \hline
 $\Lambda=380\,\MeV$ &  0.30 & -0.074(4)  &  0.025(2) &    0       \\ \hline
                     &  0.27 & -0.049(5)  & -0.007(5) &  0.009(1) \\ \hline\hline
 NLO ``Denom''       &  0.47 & -0.046(2)  &     0      &    0       \\ \hline
 $\Lambda=340\,\MeV$ &  0.17 & -0.066(2)  &  0.013(4) &    0       \\ \hline
                     &  0.17 & -0.061(4)  &  0.009(3) &  0.0005(3) \\ \hline\hline
 N$^2$LO APT         &  0.47 & -0.054(1)  &     0      &    0       \\ \hline
 $\Lambda=380\,\MeV$ &  0.17 & -0.065(2)  &  0.0081(5) &    0       \\ \hline
                     &  0.10 & -0.069(2)  &  0.0114(9) & -0.0006(1) \\ \hline
\end{tabular} \end{center}
\end{table}

 It is worth noting that the best APT fit allows one to describe well
 all the BSR data at scales down to $Q \sim  350$ MeV with only the
 first three terms of the OPE series, unlike the usual PT case, where
 such fits happened to be
 impossible (due to the ghost issue) even for an increasing number of HT
 terms. This means, that the lower bound of the pQCD applicability
 (supported by power HT terms) now may be shifted down to
 $Q\sim\Lambda_{\rm QCD}\simeq 350$ MeV.

  However, it seems to be difficult to get a description in the region
 $Q<\Lambda_{\rm QCD} $. This is not surprising, because the expansion
 in positive powers of $Q^2$ and its matching \cite{SofTer} with the HT
 expansion are relevant here. In
 this respect, $\Lambda_{\rm QCD} $ scale appears as a natural border
 between ``higher twist" and ``chiral'' nonperturbative physics.

\section{Conclusion and Outlook}

 The separation of perturbative and NP physics may be different if some
 modification of perturbation theory is adopted. To test such a separation,
 we performed a systematic comparison of HT terms extracted from the
 very accurate JLab data on BSR in the framework of both the common PT
 and yhe APT in QCD and came to the following results.
\begin{itemize}
\item  The evidence that the "Denominator" form (\ref{Denom}) of the QCD
 coupling \as is more suitable in the low $Q$ region is given (see Fig. 1).

 \item A kind of duality between HO of PT and HT is observed so that HO
 terms absorb part of HT contributions moving the pQCD frontier between
 the PT and HT contribution to lower $Q$ values (see Fig. 2).
 \item
 This situation is more pronounced in the APT where convergence of both
 the HO and HT series is much better. While the twist-4 term happened to
 be larger in magnitude in the APT than in the PT, the subsequent terms
 are essentially smaller and quickly decreasing (as the APT absorbs some
 part of NP dynamics described by HT). This is the second reason of
 the shift of pQCD frontier to lower $Q$ values.

 As the main result, a satisfactory description of the data down to
 $Q\sim\Lambda_{QCD}\simeq 350\,\MeV\,$ is achieved by taking the analytic
 HO and HT contributions into account simultaneously (see Fig. 3).
\end{itemize}
  In a sense, this could be natural if the main reason of such a success
 was the disappearance of unphysical singularities. We have in mind that
 the singularity-free APT and SGF QCD couplings are very close in the
 domain $Q\gtrsim 400\,\MeV$. Moreover, various lattice data
 \cite{latt} (see also reviews \cite{lattRev} and references therein)
 yield similar $\as$ curves there.

 It will be very interesting to explore the interplay between PT
 and NP physics against other low energy experimental data.

\vskip 0.1cm
 {\bf Acknowledgements} This work was partially supported by RFBR grants
 08-01-00686, 06-02-16215, 07-02-91557 and
08-02-00896-a, the JINR-Belorussian Grant (contract F08D-001) and RF
 Scientific School grant 1027.2008.2. We are thankful to
 A.P. Bakulev, J.P. Chen, G. Dodge, S.B. Gerasimov, A.L. Kataev,
 S.V. Mikhailov, A.V. Sidorov, O.P. Solovtsova and D.B. Stamenov for valuable discussion
 and particuliarly to A. Deur who also provided us with the
 last CLAS data. We are indebted to A.V. Radyushkin
 for careful reading of the manuscript and helpful advices.


\end{document}